\documentclass[12pt,psfig]{article}
\usepackage{epsf,graphics}
\usepackage[figuresright]{rotating}
\def\be{\begin{equation}}
\def\ee{\end{equation}}
\def\bea{\begin{eqnarray}}
\def\eea{\end{eqnarray}}

\def\gsim{\:\raisebox{-0.5ex}{$\stackrel{\textstyle>}{\sim}$}\:} 
\begin{document}
\begin{flushright}
TIFR/TH/01-34 \\
SINP/TNP/01-20
\end{flushright}
\bigskip
\begin{center}
{\bf Energy Independent Solution to the Solar Neutrino Anomaly
including the SNO Data} \\[2cm]
Sandhya Choubey$^a$, Srubabati Goswami$^a$, D.P. Roy$^b$ \\[1cm]
$^a$ Saha Institute of Nuclear Physics, \\
1/AF, Bidhannagar, Kolkata 700 064, INDIA \\
$^b$ Tata Institute of Fundamental Research \\
Homi Bhabha Road, Mumbai 400 005, INDIA
\end{center}

\vspace{1cm}

\begin{center}
{\bf Abstract}
\end{center}
\bigskip

The global data on solar neutrino rates and spectrum, including the
SNO charged current rate, can be explained by LMA, LOW or the energy
independent solution -- corresponding to near-maximal mixing.  All the
three favour a mild upward renormalisation of the Cl rate.  A mild
downward shift of the $B$ neutrino flux is favoured by the energy
independent and to a lesser extent the LOW solution, but not by LMA.
Comparison with the ratio of SK elastic and SNO charged current
scattering rates favours the LMA over the other two solutions, but by
no more than $1.5\sigma$.

\newpage

The suppression of the solar neutrino flux has been confirmed now by a
number of experiments [1-4], covering a neutrino energy range of $0.2
- 20$ MeV.  However there is still considerable uncertainty regarding
its energy dependence.  An energy independent suppression factor has
been advocated by several groups over the years [5].  This was shown
to be disfavoured however by the combined data on the total
suppression rates [6].  More recently the weight of experimental
evidence changed in favour of an energy independent solution following
the SK data on the day/night spectrum, showing practically no energy
dependence nor any perceptible day/night effect [3].

It was recently shown by us in [7] that an energy independent solution
can describe the global data on rates and spectrum satisfactorily,
with reasonable readjustments of the Cl experiment rate and the $^8B$
neutrino flux.  This has been corroborated now by other groups
[8,9].  It was also suggested in [7] that the measurement of the
charged to neutral current ratio by the SNO experiment will be able to
distinguish the energy independent solution from the conventional MSW
solutions to the rates and spectrum data, i.e. the so called LMA and
LOW solutions.  Recently the SNO experiment has produced its first
data [4], providing the charged current scattering rate with
reasonable precission.  In the absence of the neutral current data one
can combine the $CC$ rate of SNO with the elastic scattering rate of
SK to distinguish the above solutions from one another.  The purpose
of the present exercise is two fold.  Firstly we reevaluate the
experimental status of the energy independent solution vis a vis the
LMA and LOW solutions after the inclusion of the SNO data.  Secondly
we plug an important gap in the analysis of [7], which arose out of
neglecting the coherent term in the solar neutrino oscillation.
Inclusion of this term allows us to probe the parameter space down to
$\Delta m^2 \simeq 10^{-11} \ {\rm eV}^2$, i.e. the vacuum oscillation
region, which was not possible in [7].  We also take this opportunity
to update the result using the full SK data sample of 1258 days, which
became available more recently [3].  We shall see that the conclusion
of [7] are not changed by this and the inclusion of the SNO $CC$ rate
in the global rates and spectrum data.  On the other hand a direct
comparison with the ratio of SNO $(CC)$ and SK $(ES)$ scattering rates
does favour the LMA solution over the energy independent (and to a
lesser extent the LOW) solution, but at only $\sim 1.5\sigma$ level.

Table 1 shows the suppression rate or survival probability of the
solar neutrino $(P_{\nu_e \nu_e})$ from the combined Ga [1], Cl
[2], SK [3] and SNO [4] experiments along with their threshold
energies.  The corresponding compositions of the solar neutrino flux
are also indicated.  The SK survival rate shown in parantheses is
appropriate for the oscillation of $\nu_e$ into another active flavour
$(\nu_{\mu,\tau})$, on which we shall concentrate here.  It is
obtained by subtracting the neutral current contribution of
$\nu_{\mu,\tau}$ from the SK rate.  All the survival rates are shown
relative to the standard solar model (SSM)  prediction of BPB00 [10].

The apparent energy dependence in the survival rates of Table 1 is
conventionally explained in terms of the vacuum oscillation (VO),
small and large angle MSW (SMA and LMA) as well as the LOW solutions
[11].  The VO and SMA solutions, showing strong and nonmonotonic
energy dependence, are essentially ruled out now by the SK day/night
spectrum [3].  On the other hand the spectrum data is compatible with
the LMA and LOW solutions, which predict modest and monotonic decrease
of the survival rate with energy.

We shall fit the combined data on the survival rates and the SK
spectrum with the energy independent solution as well as the four
traditional solutions mentioned above.  In order to reconcile the
energy independence of the spectrum with the apparent energy
dependence in the rates of Table 1, we shall consider the following
variations in the Cl rate and $B$ neutrino flux, since the Cl
experiment [2] has not been calibrated while the $B$ neutrino flux is
very sensitive to the underlying solar model.

\begin{enumerate}
\item[{i)}] We shall consider an upward renormalisation of the Cl
rate by 20\% (i.e. $2\sigma$), which will push it marginally above the
SK and SNO rates.  This is favoured not only by the energy independent
solution but also the LMA and LOW solutions, showing monotonic energy
dependence. 

\item[{ii)}] We shall consider a downward variation of the $B$
neutrino flux of BPB00 [10],
\be
f_B = 5.15 \times 10^6/{\rm cm}^2/{\rm sec} \left(1.0 \matrix{+.20 \cr
\cr -.16}\right)
\label{one}
\ee
by upto $2\sigma$.  A downward renormalisation of this flux is
favoured by the energy independent solution and to a lesser extent by
LOW, though not by the LMA solution.  It is also favoured by 
some helioseismic models, e.g. the model of [12] giving $f_B = (4.16
\pm 0.76) \times 10^6/{\rm cm}^2/{\rm sec}$.
\end{enumerate}

The definition of $\chi^2$ used in our fits is
\be
\chi^2 = \sum_{i,j} (F_i^{\rm th} - F_i^{\rm exp}) (\sigma_{ij}^{-2})
(F_j^{\rm th} - F_j^{\rm exp}),
\label{two}
\ee
where $i,j$ run over the number of data points, and both the
experimental and theoretical values of the fitted quantities are
normalised relative to the BPB00 [10] prediction.  The experimental
values for the total rates are the ones shown in Table 1, while the SK
day/night spectra are taken from [3].  The error matrix $\sigma_{ij}$
contains the experimental errors as well as the theoretical errors of
the BPB00 predictions along with their correlations.  The latter
includes the uncertainty in the $B$ neutrino flux of eq. (\ref{one}).
The error matrix is evaluated using the procedure of [13].  The
details of the solar code used is described in [14].  As in [7] we
vary the normalisation of the SK spectrum as a free parameter to avoid
double counting with the SK data on total rate.  Thus there are $(2
\times 19 - 1)$ independent data points from the SK day/night spectrum
along with the 4 total rates giving a total of 41 points.  In addition
to the best-fit parameter values and $\chi^2_{\rm min}$ we shall
present the goodness of fit (g.o.f.) of a solution, which represents
the probability of a correct model having a $\chi^2 \geq$ this
$\chi^2_{\rm min}$.  Finally for the oscillation solutions with matter
effects we shall also delineate the 90\%, 95\%, 99\% and 99.73\%
$(3\sigma)$ allowed regions in the mass and mixing parameter $(\Delta
m^2 - \tan^2 \theta)$ plane.  These regions correspond to $\chi^2 \leq
\chi^2_{\rm min} + \Delta \chi^2$, where $\Delta \chi^2 = 4.61, 5.99,
9.21$ and $11.83$ respectively for two parameters and 
$\chi^2_{\rm min}$ is the global $\chi^2$ minimum.
\bigskip

\noindent {\bf Energy Independent Solution:}
\medskip

\nobreak
Table 2 summarises the results of fitting the global rate + spectrum
data with an energy independent survival probability
\be
P_{\nu_e \nu_e} = 1 - {1 \over 2} \sin^2 2\theta.
\label{three}
\ee
It shows that even without any readjustment to the Cl rate or the
$B$ neutrino flux the energy independent solution gives an acceptable
g.o.f. of 24\%.  An upward renormalisation of the Cl rate by 20\%
improves the g.o.f. to 42\%.  And varying the $B$ neutrino flux
downwards improves it further to 49\%, which corresponds to a
renormalisation factor $X_B = 0.7$ for the $B$ neutrino flux.  Note
however that the g.o.f. and the best-fit value of the mixing angle for
$X_B = 1$ are very close to those for $X_B = 0.7$.  This is because
the large error in the $B$ neutrino flux of eq. (\ref{one}) is already
incorporated into the error matrix of eq. (\ref{two}).  It
automatically chooses a lower value of flux, close to $X_B = 0.7$,
even without floating this parameter.

Traditionally the energy independent solution (\ref{three}) is
associated with the averaged vacuum oscillation probability at
distances much larger than the oscillation wave-length, as originally
suggested by Gribov and Pontecorvo [15].  As we shall see below
however an effectively energy independent solution holds around the
maximal mixing region over a wide range of $\Delta m^2$ even after
including all the matter effects.
\bigskip

\noindent {\bf Region of Energy Independent Solution in the $\Delta m^2 -
\tan^2 \theta$ Plane:}
\medskip

The energy dependence of the survival probability arises from
different sources in different regions of the parameter space.

\begin{enumerate}
\item[{i)}] For $\Delta m^2/E < 10^{-14} \ {\rm eV}$ the earth
regeneration effect can be safely neglected.  Then the survival
probability 
\be
P_{\nu_e \nu_e} = P_1 \cos^2 \theta + P_2 \sin^2 \theta + 2 \sqrt{P_1
P_2} \sin\theta \cos\theta \left({\Delta m^2 L \over E}\right),
\label{four}
\ee
where $L$ is the distance between the neutrino production point at the
solar core and its detection point on earth; and $P_2 \ (= 1 - P_1)$
is the probability of the produced $\nu_e$ emerging from the sun as
the heavier mass eigen-state
\be
\nu_2 = \nu_e \sin\theta + \nu_\mu \cos\theta.
\label{five}
\ee
The coherent interference term, represented by the last term of
eq. (\ref{four}), is responsible for the nonmonotonic energy
dependence of the socalled Just-So (VO) solution.

\item[{ii)}] For $\Delta m^2/E \sim 10^{-14} - 10^{-11} \ {\rm eV}$,
the coherent term is negligible while the earth regeneration
contribution can be significant.  Besides over a large part of this
region, represented by the MSW triangle, the $\nu_e$ is adiabatically
converted into $\nu_2$ in the sun, i.e. $P_2 = 1$.  Thus the day/night
averaged probability
\be
\bar P_{\nu_e \nu_e} = \sin^2\theta + {\eta_E \sin^2 2\theta \over 4(1
- 2\eta_E \cos 2\theta + \eta^2_E)},
\label{six}
\ee
where
\be
\eta_E = 0.66 \left({\Delta m^2/E \over 10^{-13} \ {\rm eV}}\right)
\left({g/{\rm cm}^2 \over \rho Y_e}\right).
\label{seven}
\ee
Here $\rho$ is the matter density in the earth and $Y_e$ the average
number of electrons per nucleon [16].  The regeneration contribution
is always positive and peaks around $\eta_E \sim 1$, i.e. $\Delta
m^2/E \sim 3 \times 10^{-13} \ {\rm eV}$.  This is responsible for the
LOW solution.

\item[{iii)}] Finally for $\Delta m^2/E > 10^{-11} \ {\rm eV}$ the
survival probability can be approximated by the average vacuum
oscillation probability of eq. (\ref{three}).  The MSW solutions (LMA
and SMA) lie on the boundary of the regions ii) and iii), i.e. $\Delta
m^2 \sim 10^{-5} \ {\rm eV}^2$ for $E \sim 1$ MeV. The survival
probability $P_{\nu_e \nu_e}$ goes down from $1 - {1\over2} \sin^2
2\theta \ (> 0.5)$ to $\sin^2\theta \ (< 0.5)$ in going up from Ga
to SK (SNO) energy.
\end{enumerate}

All the solar neutrino rates except that of SK have $\gsim 10\%$ error,
which is also true for the SK energy spectrum.  The SK normalisation
has at least similar uncertainty from the $B$ neutrino flux.
Therefore we shall 
treat solutions, which predict survival probability $P_{\nu_e \nu_e}$
within 10\% of eq. (\ref{three}) over the Ga to SK energy range, as
effectively energy independent solutions [7].  Moreover the predicted
Ga, Cl and SNO rates will be averaged over the respective energy
spectra, while the predicted SK rates will be averaged over 0.5 MeV
bins, corresponding to the SK spectral data, since experimental
information is available for only these averaged quantities.

Fig. 1 shows the region of effectively energy independent solution as
per the above definition.  The parameter space has been restricted to
$\Delta m^2 < 10^{-3} \ {\rm eV}^2$ in view of the constraint from the
CHOOZ experiment [17].  One sees that the energy independent solution
(\ref{three}) is effectively valid over the two quasi vacuum
oscillation regions lying above and below the MSW range.  Moreover it
is valid over a much larger range of $\Delta m^2$ around the maximal
mixing region, since the solar matter effect does not affect $P_{\nu_e
\nu_e}$ at $\tan^2\theta = 1$.  It is this near-maximal mixing strip
that is relevant for the observed survival probability, $P_{\nu_e
\nu_e} \sim 1/2$.  The upper strip $(\Delta m^2 = 10^{-3} - 10^{-5} \
{\rm eV}^2)$ spans the regions iii) and part of ii) till it is cut off
by the earth regeneration effect.  The lower strip $(\Delta m^2 =
10^{-7} - 5 \times 10^{-10} \ {\rm eV}^2)$ spans parts of region ii)
and i) till it is cut off by the coherent term contribution.  Note
that this near-maximal mixing strip represents a very important region
of the parameter space, which is favoured by the socalled bimaximal
mixing models of solar and atmospheric neutrino oscillations [18,19]. 

One can easily check that averaging over the SK energy bins of 0.5 MeV
has the effect of washing out the coherent term contribution for
$\Delta m^2 > 2 \times 10^{-9} \ {\rm eV}^2$.  Hence the contour of
effectively energy independent solution, shown in ref. [7] neglecting
the coherent term, is correct.  But including this term enables us now
to trace the contour down to its lower limit.  It was claimed in
ref. [9] that the lower strip disappears when one includes the
coherent term contribution.  This may be due to the fact that their
predicted rate in the SK energy range was not integrated over the
corresponding bin widths of 0.5 MeV.

To get further insight into the oscillation phenomenon in the maximal
mixing region we have plotted in Fig. 2 the predicted survival rates
at maximal mixing against $\Delta m^2$ for the Ga, Cl, SK and SNO
experiments.  In each case the rate has been averaged over the
corresponding energy spectrum.  This is similar to the Fig. 7 of
Gonzalez-Garcia, Pena-Garay, Nir and Smirnov [16].  As in [16] the
predictions have been shown relative to the central value of the $B$
neutrino flux of BPB00 along with those differing by $\pm 20\%$ from
the central value, i.e. $X_B = 1 \pm 0.2$.  We have found that these
curves are in good agreement with the corresponding ones of [16].  The
two regions of $< 10\%$ energy dependence are indicated by vertical
lines.  One can easily check that in these regions the central curves
lie within 10\% of the energy independent prediction $R = 0.5$ (only
the SK rate is higher due to the neutral current contribution).  One
can clearly see the violent oscillation in the Just-So (VO) region at
the left and the LOW solution corresponding to the earth regeneration
peak of the Ga experiment in the middle.  The LMA and SMA are not
identifiable since they do not occur at maximal mixing angle.  It should be
noted that the gap between the two energy independent regions due to
the earth regeneration effect in Fig. 1 is a little narrower than
here.  This is due a cancellation between the positive contribution
from the earth regeneration effect and the negative contribution from
the solar matter effect at $\tan^2\theta < 1$.  It ensures that the
resulting survival rate agrees with the energy independent solution
(\ref{three}) over a somewhat larger range of $\Delta m^2$ slightly
below the maximal mixing angle.

The observed rates from the Ga, Cl, SK and SNO experiments are shown
in Fig. 2 as horizontal lines along with their $1\sigma$ errors.  With
20\% downward renormalisation of the $B$ neutrino flux $(X_B = 0.8)$
the energy independent prediction is seen to agree with the SK rate and
also approximately with SNO.  It is higher than the Cl rate by about
$2\sigma$.  The agreement with the Ga rate can be improved by going
to a little smaller $\theta$ and compensating the resulting deviation
from the other rates by a somewhat smaller $X_B$ as in Table 2.
Nonetheless the maximal mixing solution for $X_B = 0.8$, shown here,
is in reasonable agreement with the observed rates over the energy
independent regions.  The earth regeneration effect can be seen to
improve the agreement with the Ga experiment for the LOW solution.
\bigskip

\noindent {\bf The SMA, LMA, LOW and VO Solutions:}
\medskip

\nobreak
Tables 3 and 4 summarises the results of fits to the global rates +
spectrum data in terms of the conventional oscillation solutions.
Table 3 shows solutions to the data with observed and renormalised
Cl rate with the neutrino flux of BPB00 $(X_B = 1)$, while Table 4
shows the effects of renormalising this $B$ neutrino flux downwards by
25\% $(X_B = 0.75)$.  The corresponding 90\%, 95\%, 99\% and 99.73\%
$(3\sigma)$ contours are shown in Fig. 3.

As we see from these tables and Fig. 3 the SMA solution gives rather
poor fit in each case, with no allowed contour at $3\sigma$ level.
This result agrees with the recent fits of [20] to the global data
including SNO.  The fit of [21] to these data shows a small allowed
region for SMA solution at $3\sigma$ level due to a slightly different
method of treating the normalisation in the SK spectrum data, as
explained there.

The LMA and LOW solutions give good fits to the original data set,
which improve further with the upward renormalisation of the Cl rate
by $2\sigma$.  This is because the monotonic decrease of rate with
energy, implied by these solutions, favours the Cl rate to be
marginally higher than the SNO and SK rates as mentioned earlier.
For the renormalised Cl case, downward renormalisation of 
the $B$ neutrino flux by 25\% is seen to give
a modest increase (decrease) of g.o.f. for the best LOW (LMA) solution.
On the other hand the allowed ranges increase in both cases as we see
from Fig. 3b and c.  Combined together they imply that the downward
renormalisation of the $B$ neutrino flux modestly favours the LOW
solution but makes little difference to the LMA.  Its main effect on
these two solutions is increasing their allowed ranges in the
parameter space.  Comparing Fig. 3c with Fig. 1 shows that much of
these enlarged ranges of validity correspond to effectively energy
independent solution.
It is intersting to note that
the best-fit values of parameters in the LMA region are 
same for Cl observed and Cl renormalised cases while the $\chi^2_{min}$ 
improves for the latter. 
This shows that the  best-fit already chose
a probability at Cl energy, which is a little higher than 
that at SK/SNO energy.
Renormalising the Cl rate  brought that point up to the fitted curve
without changing the best-fit parameters. 

While the best vacuum solution seems to show remarkably high
g.o.f. particularly for renormalised Cl rate and $B$ neutrino flux,
its regions of validity are two miniscule islands just below the lower
energy independent strip of Fig. 3b,c.  This solution has also been
obtained in the global fits of ref. [20,21] as well as the SK fit to
their rate and spectrum data [3].  The position and size of its range
of validity suggest this to be a downward fluctuation of the
effectively energy independent quasi vacuum oscillation rather than a
genuine VO solution of Just-So type.  To get further insight into this
solution we have analysed the resulting energy dependence.  It shows
practically no energy dependence below 5 MeV, but a 15\% fall over the
$5-12$ MeV range.  The latter seems to follow the SK spectral points
rather closely amidst large fluctuation.  To check the stability of
this trend we have repeated the fit to the SK spectral data points,
plotted over 8 broad energy bins shown in [3], which show much less
fluctuation than the $2 \times 19$ points sample.  The solution
completely disappears from this fit.  This confirms that the above VO
solution is simply an artifact of the sampling of the SK spectral
data.

For completeness we summarise in Table 5 the best fits of the above
solutions with free $B$ neutrino flux normalisation.  The SMA solution
favours a very low $B$ neutrino flux $(X_B \simeq 0.5)$, which raises
the SK and SNO rates more than the Cl, thus accentuating the
nonmonotonic energy dependence of Table 1.  Still the g.o.f. of the
SMA solution is rather marginal.  On the other hand the LMA solution
favours $X_B > 1$, which suppresses the SK and SNO rates more than the
Cl, resulting in a monotonic decrease of rate with energy.  But the
corresponding g.o.f. are no better than those of Table 3.  The results
of the LOW and VO fits are similar to those of Table 4.
\bigskip

\noindent {\bf Comparison with the Ratio of SK and SNO Rates:}
\medskip

\nobreak
With the measurement of both charged and neutral current scattering
rates at SNO it will be possible to discriminate between the above
solutions, since the $B$ neutrino flux factors out from their ratio
[7].  In the absence of the neutral current data from SNO one can try
to make a similar comparison with the ratio of SK elastic and SNO
charged current scattering rates, 
\be
R^{el}_{SK} = X_B P_{\nu_e \nu_e} + r(1 - P_{\nu_e \nu_e}) X_B, \ r =
\sigma_{nc}/\sigma_{cc} \simeq 0.17,
\label{eight}
\ee
\be
R^{cc}_{SNO} = X_B P_{\nu_e \nu_e},
\label{nine}
\ee
where we have assumed a common survival rate neglecting the small
difference between the SK and SNO energy spectra [22].  One can
eliminate $P_{\nu_e \nu_e}$ from the two rates; and the resulting $B$
neutrino flux can be seen to be in good agreement with the BPB00
estimate [4].  Alternatively one can factor out the flux from the
ratio 
\be
R^{es}_{SK}/R^{cc}_{SNO} = 1 - r + r/P_{\nu_e \nu_e}.
\label{ten}
\ee

Table 6 shows the best fit values of the above ratio for the LMA, LOW
and the energy independent solutions along with the corresponding
predictions for $R^{cc}_{SNO}/R^{nc}_{SNO} = P_{\nu_e \nu_e}$.  The
predictions of the maximal mixing solution is also shown for
comparison.  While the predictions for the $CC/NC$ ratio differ by
$\sim 50\%$ they differ by only about $\sim 15\%$ in the case of the
$ES/CC$ ratio.  The observed ratio $R^{es}_{SK}/R^{cc}_{SNO}$ is seen
to favour the LMA over the LOW and energy independent solutions; but
even the largest discrepancy is only $\sim 1.5\sigma$.  With the
expected sample of several thousand $CC$ and $NC$ events from SNO one
expects to reduce the $1\sigma$ error for each of these ratios to
about 5\%.  Then one will be able to discriminate between the three
solutions meaningfully, particularly with the help of the $CC/NC$
ratio from SNO.

\vskip 20pt 
S.C. and S.G. would like to acknowledge Abhijit Bandyopadhyay for 
discussions. 

\newpage

\begin{table}
\caption{
 The ratio of the observed solar neutrino rates to the
corresponding BPB00 SSM predictions.
}
\[
\begin{tabular}{ccc} \hline
experiment & $\frac{obsvd}{BPB00}$ & composition \\ \hline Cl &
0.335 $\pm$ 0.029 & $B$ (75\%), $Be$ (15\%)
\\
Ga & 0.584 $\pm$ 0.039 &$pp$ (55\%), $Be$ (25\%), $B$ (10\%)
\\
SK & 0.459 $\pm$ 0.017  & $B$ (100\%)
\\
& (0.351 $\pm$0.017) & \\ SNO(CC) & 0.347 $\pm$ 0.027 & $B$
(100\%) \\ \hline \hline
\end{tabular}
\]
\end{table}

\begin{table}[htb]
\caption{
The best-fit value of the parameter, the $\chi^2_{\min}$
and the g.o.f from a combined analysis of rate and spectrum with the
energy independent solution given by eq. (3).}
\begin{center}
\begin{tabular}{ccccc}
\hline\hline
&$X_{B}$&$\sin^22\theta\left(\matrix{\tan^2\theta \cr {\rm
or} \cr \cot^2 \theta}\right)$& $\chi^2_{\min}$ & g.o.f   \\
\hline
Chlorine& 1.0&0.93(0.57)& 46.06 &23.58\%\\
Observed&0.72 & 0.94(0.60) & 44.86 & 27.54\%\\
\hline
Chlorine&1.0 &0.87(0.47) &41.19& 41.83\%\\
Renormalised&0.70&0.88(0.48)&38.63 & 48.66\%\\
\hline \hline
\end{tabular}
\end{center}
\end{table} 

\begin{table}
\caption
{The best-fit values of the parameters, the $\chi^2_{\min}$
and the g.o.f from a combined analysis of the Cl, Ga, SK and SNO CC
rates and the SK day-night spectrum in terms
of $\nu_e$ oscillation into an active neutrino, including the
matter effects. $X_{B}$ is kept fixed at the SSM value (=1.0). }
\begin{center}
\begin{tabular}{cccccc}
\hline\hline
& Nature of & $\Delta m^2$ &
& & \\
& Solution & in eV$^2$& \raisebox{1.5ex}[0pt] {$\tan^2\theta$} &
\raisebox{1.5ex}[0pt] {$\chi^2_{min}$}&\raisebox{1.5ex}[0pt] {g.o.f}\\
\hline
& SMA & $5.28 \times 10^{-6}$&$3.75\times 10^{-4}$
&51.14 &9.22\%  \\
Cl& LMA & $4.70\times 10^{-5}$ &0.38&
33.42 & 72.18\%  \\
Obsvd.& LOW & $1.76\times 10^{-7}$ & 0.67 &39.00&46.99\%  \\
&VO&$4.64\times 10^{-10}$ & 0.57 & 38.28 & 50.25\% \\
\hline
& SMA & $4.94 \times 10^{-6}$& $2.33 \times 10^{-4}$
& 50.94& 9.54\%  \\
Cl& LMA & $4.70\times 10^{-5}$ &0.38&30.59& 82.99\%\\
Renorm.& LOW & $1.99 \times 10^{-7}$ &0.77& 34.26& 68.57\% \\
&VO&$4.61\times 10^{-10}$ & 0.59 & 32.14 & 77.36\% \\
\hline\hline
\end{tabular}
\end{center}
\end{table}

\begin{table}
\caption
{Best fits to the combined rates and SK day-night spectrum data in
terms of $\nu_e$ oscillation into active neutrino with a
fixed $X_B = 0.75$.}
\begin{center}
\begin{tabular}{cccccc}
\hline\hline
& Nature of & $\Delta m^2$ &
& & \\
& Solution & in eV$^2$& \raisebox{1.5ex}[0pt] {$\tan^2\theta$} &
\raisebox{1.5ex}[0pt] {$\chi^2_{min}$}&\raisebox{1.5ex}[0pt] {g.o.f}\\
\hline
& SMA & $5.28 \times 10^{-6}$&$3.75\times 10^{-4}$
&48.39 &14.40\%  \\
Cl& LMA & $4.65\times 10^{-5}$ &0.49&
38.90 & 47.44\%  \\
Obsvd.& LOW & $1.74\times 10^{-7}$ & 0.71 &39.91&42.95\%  \\
& VO & $4.55\times 10^{-10}$ & 0.44 & 37.17 & 55.36\%\\
\hline
& SMA & $8.49 \times 10^{-6}$& $1.78 \times 10^{-4}$
& 50.77& 9.82\%  \\
Cl& LMA & $4.64\times 10^{-5}$ &0.51&34.48& 67.61\%\\
enorm.&LOW & $2.09 \times 10^{-7}$ &0.81& 33.47& 71.97\% \\
&VO & $4.59\times 10^{-10}$&0.53&30.63& 82.86\%\\
\hline\hline
\end{tabular}
\end{center}
\end{table}

\begin{table}
\caption
{Best fits to the combined rates and SK day-night spectrum data in
terms of $\nu_e$ oscillation into active neutrino with 
$X_B$ free.}
\begin{center}
\begin{tabular}{ccccccc}
\hline
& Nature of & &$\Delta m^2$ &
& & \\
& Solution & \raisebox{1.5ex}[0pt] {$X_B$}&
in eV$^2$& \raisebox{1.5ex}[0pt] {$\tan^2\theta$} &
\raisebox{1.5ex}[0pt] {$\chi^2_{min}$}&\raisebox{1.5ex}[0pt] {g.o.f}\\
\hline
&SMA & 0.51& $5.25 \times 10^{-6}$&$3.44\times 10^{-4}$
&46.83 &15.41\%  \\
Cl& LMA & 1.18& $4.73\times 10^{-5}$ &0.33&
32.32 & 72.89\%  \\
Obsvd.& LOW &0.88&  $1.75\times 10^{-7}$ & 0.67 &38.75&43.57\%  \\
&VO& 0.70&$4.55\times 10^{-10}$& 0.44 & 37.24 & 50.44\%\\
\hline
& SMA & 0.48 & $4.66 \times 10^{-6}$& $2.32 \times 10^{-4}$
& 46.18& 17.01\%  \\
Cl& LMA & 1.15 & $4.71\times 10^{-5}$ &0.36&30.32& 80.80\%\\
Renorm.&LOW & 0.83 & $2.03 \times 10^{-7}$ &0.79& 33.18& 69.17\% \\
&VO&0.75&$4.63\times 10^{-10}$ & 0.55& 30.56 & 79.92\%\\
\hline\hline
\end{tabular}
\end{center}
\end{table}

\begin{table}
\caption{The values of the ratios $R^{ES}_{SK}/R^{CC}_{SNO}$ and 
$R^{CC}_{SNO}/R^{NC}_{SNO}$ at the best-fit values for the 
LMA,LOW and energy independent  solutions for the  
renormalised Cl and $X_{B}$ =1.0 case. 
Also shown are the predictions for the  maximal mixing ($P_{ee}=0.5$) 
solution and  the experimental value of 
$R^{ES}_{SK}/R^{CC}_{SNO}$.} 

\begin{center}
\begin{tabular}{cccccc}
\hline
& $\Delta m^2$ & $\tan^2\theta$ &  $R^{CC}_{SNO}/R^{NC}_{SNO}$ &
$R^{ES}_{SK}/R^{CC}_{SNO}$ & Expt. value of \\ 
& & & & & $R^{ES}_{SK}/R^{CC}_{SNO}$ \\ \hline 
LMA & $4.7 \times 10^{-5}$ & 0.38 & 0.30 & 1.36 & \\
LOW & $1.99 \times 10^{-7}$ & 0.77 & 0.45 & 1.19 &  \\
energy-independent & - & 0.47 & 0.56 & 1.13 &  \raisebox{1.2ex}[0pt]{1.33 $\pm$ 0.13} \\
maximal mixing & - & 1.0 & 0.5 & 1.15 & \\
\hline\hline
\end{tabular}
\end{center}
\end{table}

\newpage

\begin{center}
{\bf References}
\end{center}
\bigskip

\begin{enumerate}

\item[{1.}] GNO Collaboration: M. Altmann et. al., Phys. Lett. B490,
16 (2000); Gallex Collaboration: W. Hampel et. al., Phys. Lett. B447,
127 (1999); SAGE Collaboration: J.N. Abdurashitov et. al.,
Phys. Rev. C60, 055801 (1999).

\item[{2.}] B.T. Cleveland et. al., Astrophys. J. 496, 505 (1998).

\item[{3.}] SK Collaboration: S. Fukuda et. al., hep-ex/0103033.

\item[{4.}] SNO Collaboration: Q.R. Ahmad et. al., nucl-ex/0106015.

\item[{5.}] A. Acker, S. Pakvasa, J. Learned and T.J. Weiler,
Phys. Lett. B298, 149 (1993); P.F. Harrison, D.H. Perkins and
W.G. Scott, Phys. Lett. B349, 137 (1995) and B374, 111 (1996); R. Foot
and R.R. Volkas, hep-ph/9570312; A. Acker and S. Pakvasa,
Phys. Lett. B397, 209 (1997); G. Conforto et. al., Phys. Lett. B427,
314 (1998); W.G. Scott, hep-ph/0010335.

\item[{6.}] P.I. Krastev and S.T. Petkov, Phys. Lett. B395, 69 (1997).

\item[{7.}] S. Choubey, S. Goswami, N. Gupta and D.P. Roy,
Phys. Rev. D64, 053002 (2001).

\item[{8.}] G.L. Fogli, E. Lisi and Palazzo, hep-ph/0105080.

\item[{9.}] V. Berezinsky, M.C. Gonzalez-Garcia and C. Pena-Garay,
hep-ph/0105294. 

\item[{10.}] J.N. Bahcall, M.H. Pinsonneault and Sarbani Basu,
astro-ph/0010346. 

\item[{11.}] J.N. Bahcall, P.I. Krastev and A.Y. Smirnov,
Phys. Rev. D58, 096016 (1998). 

\item[{12.}] H.M. Antia and S.M. Chitre, A \& A, 339, 239 (1998);
S. Choubey, S. Goswami, K. Kar, H.M. Antia and S.M. Chitre,
hep-ph/0106168, Phys. Rev. D (in press).

\item[{13.}] G.L. Fogli and E. Lisi, Astropart. Phys. 3, 185 (1995).

\item[{14.}] S. Goswami, D. Majumdar and A. Raychaudhuri,
hep-ph/9909453; Phys. Rev. D63, 013003 (2001); A. Bandyopadhyay,
S. Choubey and S. Goswami, hep-ph/0101273, Phy. Rev. {\bf D63},
113019, (2001) .

\item[{15.}] V.N. Gribov and B. Pontecorvo, Phys. Lett. B28, 493
(1969). 

\item[{16.}] M.C. Gonzalez-Garcia, C. Pena-Garay, Y. Nir and
A.Y. Smirnov, Phys. Rev. D63, 013007 (2001).

\item[{17.}] M. Apollonio et. al., Phys. Lett. B446, 415 (1999).

\item[{18.}] V. Barger, S. Pakvasa, T.J. Weiler and K. Whisnant,
Phys. Lett. B437, 107 (1998).

\item[{19.}] For theoretical models of bimaximal neutrino mixing see
e.g. Y. Nomura and T. Yanagida, Phys. Rev. D59, 017303 (1999);
G. Altarelli and F. Feruglio, Phys. Lett. B439, 112 (1998); E. Ma,
Phys. Lett. B442, 238 (1998); R.N. Mohapatra and S. Nussinov,
Phys. Lett. B441, 299 (1998); R. Barbieri, L.J. Hall and A. Strumia,
Phys. Lett. B445, 407 (1999); H. Georgi and S.L. Glashow,
Phys. Rev. D61, 097301 (2000); S.F. King and G.G. Ross,
hep-ph/0108112; J. Pantaleone, T.K. Kuo and G.H. Wu, hep-ph/0108137.

\item[{20.}] G.L. Fogli, E. Lisi, D. Montanino and A. Palazzo,
hep-ph/0106247; A. Bandyopadhyay, S. Choubey, S. Goswami and K. Kar,
hep-ph/0106264 (to appear in Phys. Lett. B). 

\item[{21.}] J.N. Bahcall, M.C. Gonzalez-Garcia and C. Pana-Garay,
hep-ph/0106258. 

\item[{22.}] V. Barger, D. Marfatia and K. Whisnant, hep-ph/0106207. 

\end{enumerate}

\begin{figure}
\centerline{\epsfxsize=1.0\textwidth\epsfbox{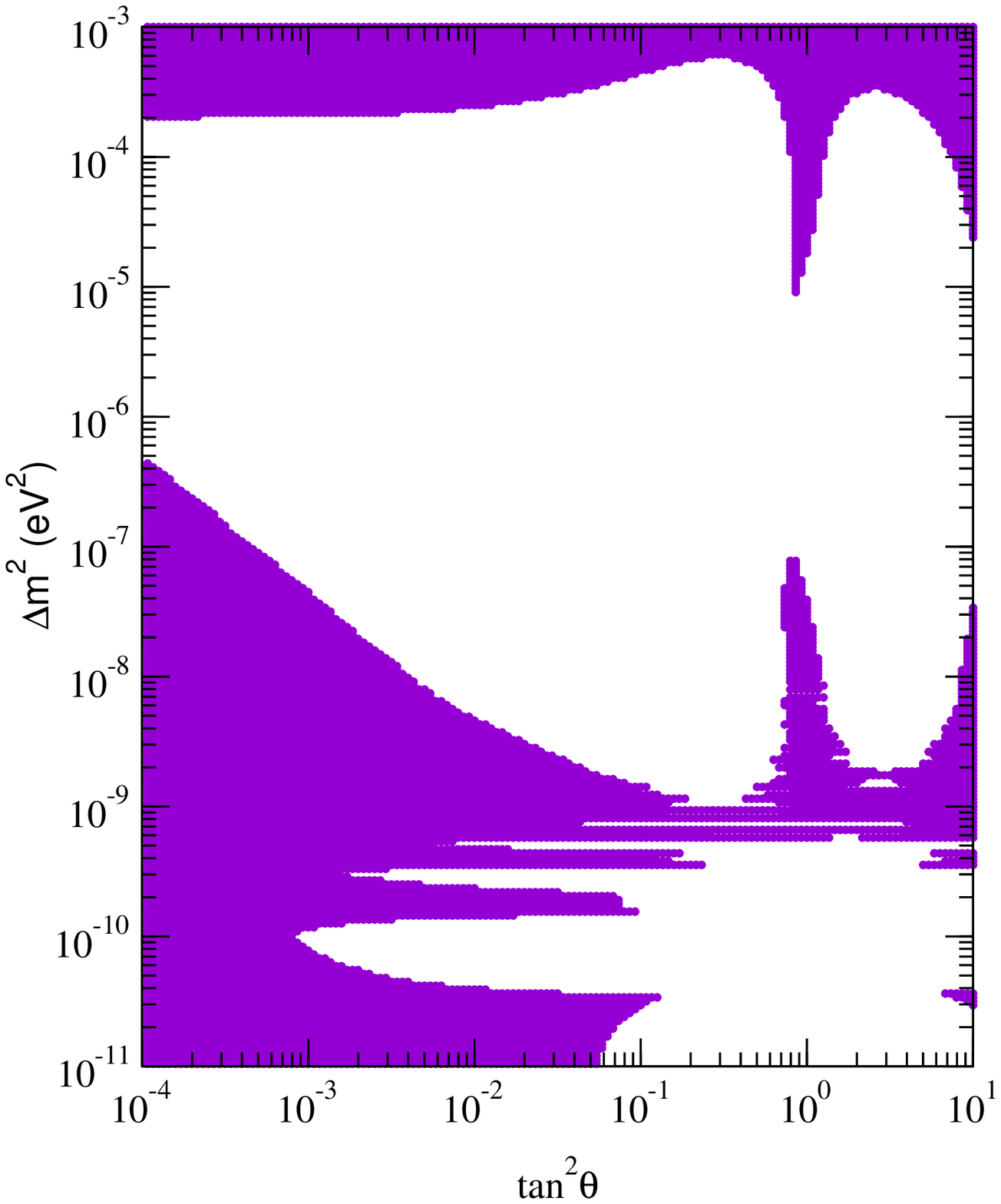}}
\vskip -1in
\parbox{5in}{
{\bf Fig 1:} The region of effectively energy independent
solution in $\Delta m^2-\tan^2 \theta$ parameter
space where the solar neutrino survival probability agrees with
eq. (3) to within 10\% over the range of Ga to SK energies.}
\end{figure}

\newpage
\begin{figure}
\topmargin -1in
\centerline{\epsfxsize=0.9\textwidth\epsfbox{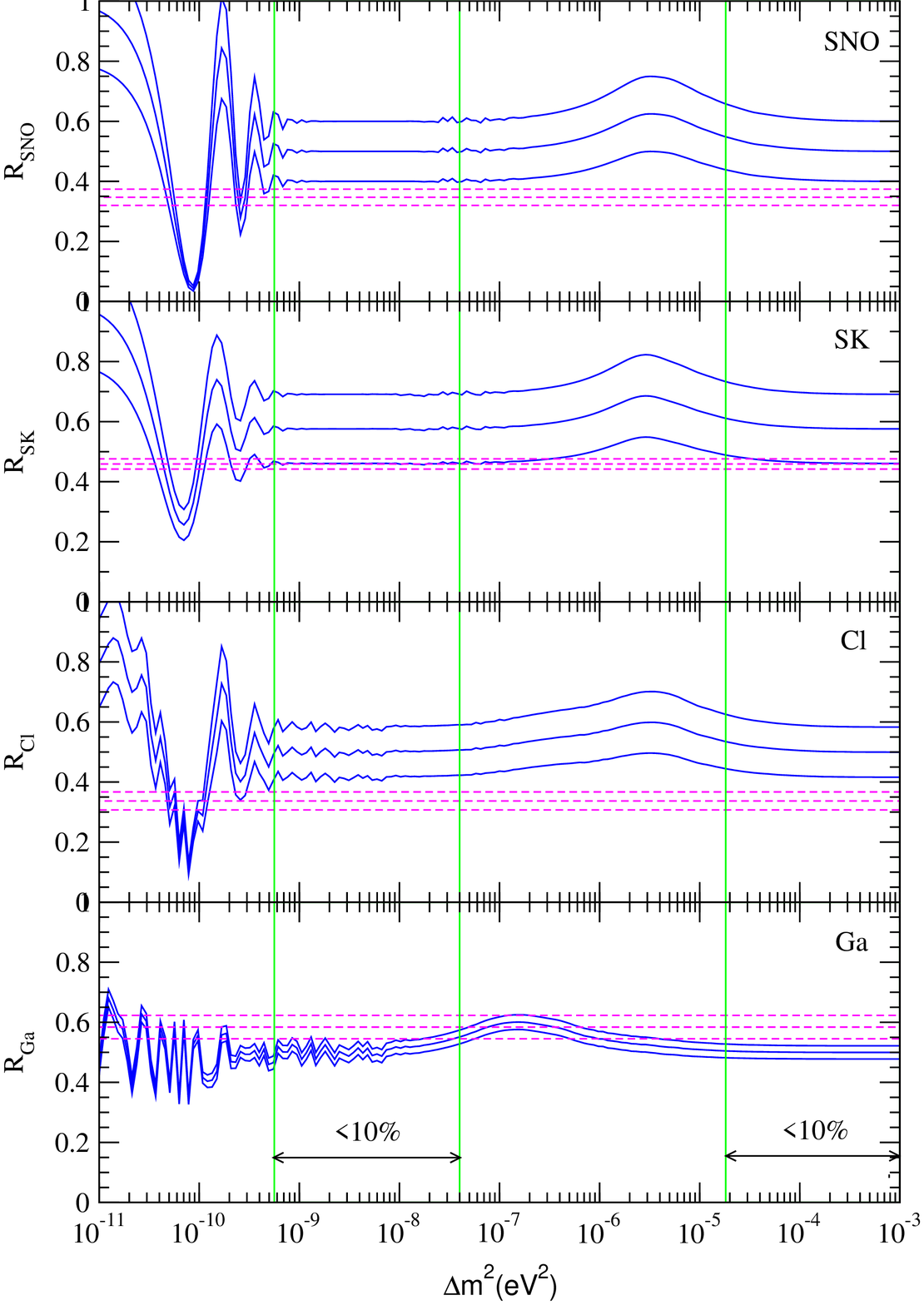}}
\topmargin -1in
\vskip 0in
\centerline{\parbox{5in}{
{\bf Fig. 2}: The predicted survival rates at maximal mixing 
against $\Delta m^2$ for Cl, Ga, SK and SNO experiments (See text for details.)
}}
\end{figure}

\newpage
\begin{figure}
\topmargin -1in
\centerline{\epsfxsize=1.0\textwidth\rotatebox{270}{\epsfbox{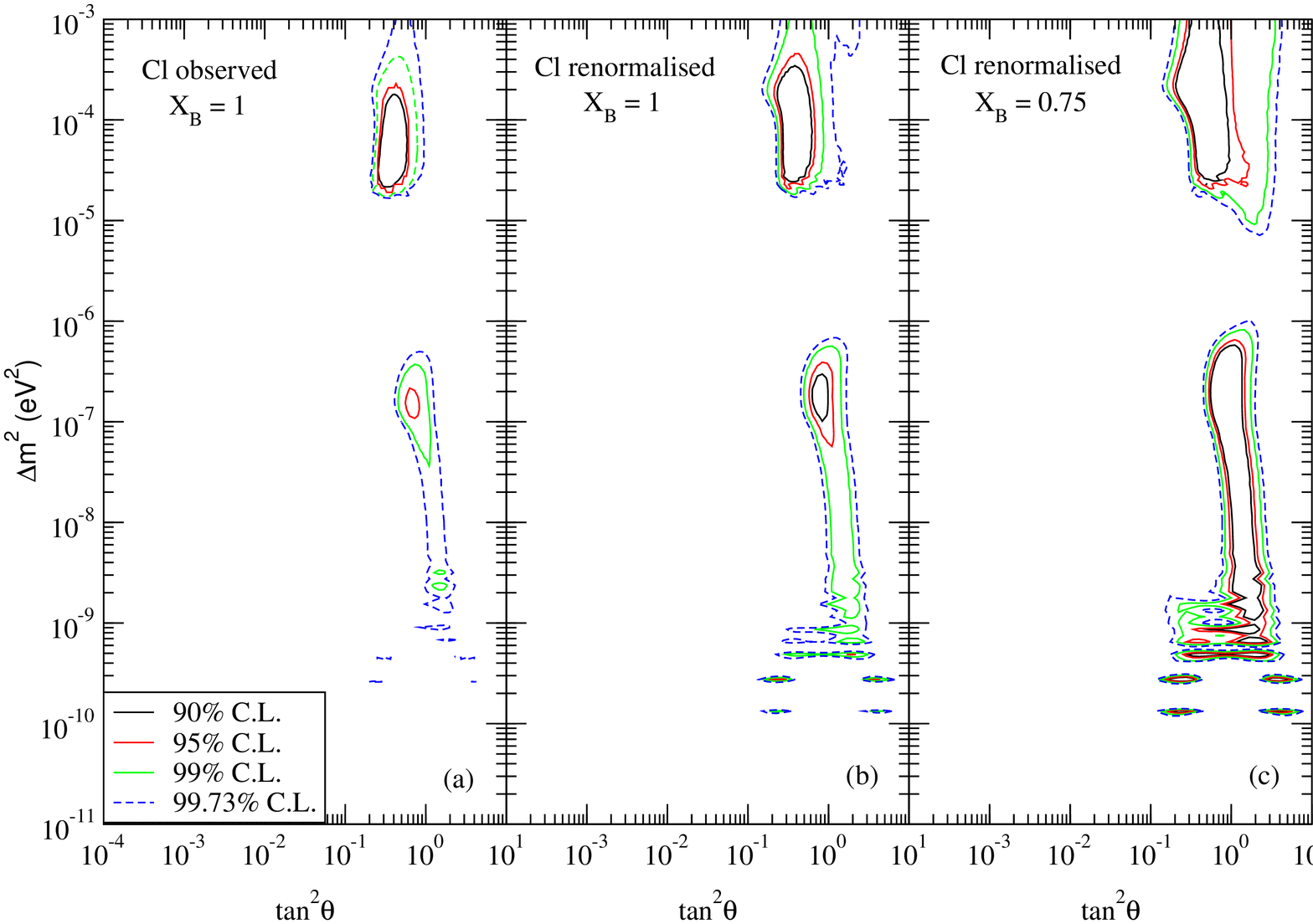}}}
\centerline{\parbox{7in}{
{\bf Fig. 3}: The 90, 95 and 99 and 99.73\% C.L. allowed area from the
global analysis of the total rates from Cl (observed  and
20\% renormalised), Ga,SK and SNO (CC) detectors
and the 1258 days SK recoil electron spectrum at day and night,
assuming MSW conversions to active neutrinos.}} 
\end{figure}
 
\end{document}